\documentstyle[sprocl,epsfig,picins]{article}

\def\ee{e^+e^-}
\def\qq{q\bar q}

\def\HW{\textsc{\small HERWIG}}
\def\pythia{\textsc{\small PYTHIA}}
\def\hpp{\mbox{\textsf{Herwig++}}}

\def\Pythia7{\textsf{Pythia7}}
\def\SHERPA{\textsf{SHERPA}}

\def\njet{\langle n_{\rm jet}\rangle}

\begin{document}

\title{\hpp{} for \boldmath $\ee$ collisions
  \footnote{Presented at the \emph{International Conference on Linear
  Colliders}, 19--23 April 2004, Paris, France (LCWS 2004).}}
\author{Stefan Gieseke}
\address{University of Cambridge, Cavendish Laboratory,\\
  Madingley Road, Cambridge CB3\,0HE, United Kingdom}

\maketitle\abstracts{Some results obtained with the new Monte Carlo
  event generator \hpp{} are presented.  In its first version (1.0),
  \hpp{} is capable of simulating $\ee$ Annihilation events.  We
  discuss its relevance for future $\ee$ colliders and show results on
  the multiplicity of minijets.}
  
\piccaption[]{$\njet$ in the $k_\perp$ algorithm.\label{fig}} 
\parpic(5.3cm,7.7cm)[r]{
  \epsfig{file=njets_NLC_kt.mpps,scale=0.6} } 
\piccaption[]{}
The advent of future colliders will require new tools for the Monte
Carlo (MC) simulation of final states.  As the well--established
tools, e.g.\ \pythia{} \cite{Pythia} and \HW{} \cite{Herwig64} have
grown tremendously during the LEP era and are considered difficult to
maintain and even more difficult to extend, it was decided to rewrite
them in an object oriented language, in particular C++.  Therefore,
the old codes are currently being rewritten under the names \Pythia7{}
\cite{Pythia7} and \hpp{} \cite{Herwig++}.  Furthermore, a new project
with the same goals, \SHERPA{} \cite{SHERPA}, has been established.
In this talk we consider our own project \hpp{}.  This program is
available in version 1.0 and capable of simulating $\ee$ annihilation
events, particularly at LEP energies.  

The main steps of the physical simulation of an $\ee$ annihilation
consist of a hard subprocess, a parton shower evolution, cluster
hadronization and hadronic decays.  In \hpp{}~1.0 we have a basic
$\ee\to(\gamma, Z)\to\qq$ tree level matrix element (ME).  In
addition, we apply hard and soft ME corrections in order to be able to
match up the phase space accessible by the parton shower with that of
the full matrix element at the lowest order for $\qq g$ production
\cite{MikeMEC}. We carry out the parton shower evolution in terms of a
set of new evolution variables as described in great detail in
\cite{NewVariables}.  These allow a smooth coverage of soft gluon
phase space and a correct treatment of heavy quarks in the quasi
collinear limit.  Some modifications of the hadronization model have
been made in order to improve baryon multiplicity predictions.

We used the program \hpp{} to simulate a large number of observables
in $\ee$ collisions and compared the results to LEP and SLD data as
well as older data from PETRA.  We covered the available spectrum of
data ranging from exclusive hadron multiplicities and momentum
distributions over all kinds of event shape distribution to four jet
angles.  We wanted to give a reasonable overall description of the
available data without emphasising any particular observable too much
and came up with a 'recommended' parameter set for \hpp{}
\cite{Herwig++}.  In Fig.~\ref{fig} we show the minijet multiplicity
in the $k_\perp$ algorithm for a future linear collider as predicted
by \hpp{}.

In conclusion, we have succeeded in delivering the new Monte Carlo
event generator \hpp{} which is perfectly capable of describing LEP
physics with a similar or better quality as \HW{} did.  This generator
will be perfectly suitable to simulate $\ee$ annihilation events at a
future linear collider.  The simulation of hadronic events will be
possible in future versions.


\begin{thebibliography}{99}
\bibitem{Herwig64} 
  G.~Corcella {\it et al.},
  ``HERWIG 6.5 release note,''
  arXiv:hep-ph/0210213; \\
  G.~Corcella {\it et al.},
  JHEP {\bf 0101} (2001) 010
  [arXiv:hep-ph/0011363]. 

\bibitem{Pythia} 
  T.~Sjostrand, L.~Lonnblad, S.~Mrenna and P.~Skands,
  arXiv:hep-ph/0308153; 
  T.~Sjostrand, P.~Eden, C.~Friberg, L.~Lonnblad, G.~Miu, S.~Mrenna and E.~Norrbin,
  Comput.\ Phys.\ Commun.\  {\bf 135} (2001) 238
  [arXiv:hep-ph/0010017].
  
\bibitem{Pythia7} 
  M.~Bertini, L.~Lonnblad and T.~Sjostrand,
  Comput.\ Phys.\ Commun.\  {\bf 134} (2001) 365
  [arXiv:hep-ph/0006152]. 

\bibitem{SHERPA}
T.~Gleisberg, S.~Hoche, F.~Krauss, A.~Schalicke, S.~Schumann and J.~C.~Winter,
JHEP {\bf 0402} (2004) 056
[arXiv:hep-ph/0311263].
 
\bibitem{Herwig++}
S.~Gieseke, A.~Ribon, M.~H.~Seymour, P.~Stephens and B.~Webber,
JHEP {\bf 0402}, 005 (2004)
[arXiv:hep-ph/0311208].


\bibitem{NewVariables}
S.~Gieseke, P.~Stephens and B.~Webber,
JHEP {\bf 0312}, 045 (2003)
[arXiv:hep-ph/0310083].

\bibitem{MikeMEC} 
M.~H.~Seymour,
Comput.\ Phys.\ Commun.\  {\bf 90} (1995) 95
[arXiv:hep-ph/9410414].

\bibitem{kupco}
  A.~Kupco,
  in Proc.\ \emph{Monte Carlo generators for HERA physics},
  (Hamburg 1998--1999),  eds.\ A.T. Doyle, G. Grindhammer, G. Ingelman, H. Jung. Hamburg, DESY-PROC-1999-02, 292 [arXiv:hep-ph/9906412].  

\end{thebibliography}
\end{document}